\documentclass[nonacm, sigconf]{acmart}
\AtBeginDocument{%
  }

\setcopyright{none}
\copyrightyear{2026}
\acmYear{2026}
\acmDOI{XXXXXXX.XXXXXXX}
\acmConference[CHIWORK'26]{the CHIWORK Workshop on Human Agency and Skill in AI‑Supported Work}{June 22,
  2026}{Linz, Austria}

\AtBeginDocument{%
  \fancypagestyle{firstpagestyle}{%
    \fancyhf{}
    \fancyhead[L]{%
      \parbox{\dimexpr\textwidth-2em\relax}{\footnotesize
      Workshop paper presented at the Workshop on Human Agency and Skill in AI-Supported Work, held in conjunction with the 5th Annual Symposium on Human-Computer Interaction for Work (CHIWORK'26) in Linz, Austria, on June 22, 2026.
      }%
    }%
    \fancyhead[R]{}%
  }
}

\begin{document}

\title{Open Questions Towards Skill-Sustaining Reliance in Reflective AI Engagement}

\author{Sander de Jong}
\orcid{0000-0002-2591-3805}
\affiliation{%
  \institution{Aalborg University}
  \city{Aalborg}
  \country{Denmark}
}
\email{sanderdj@create.aau.dk}



\renewcommand{\shortauthors}{S. de Jong}

\begin{abstract}
As AI systems are increasingly integrated into professional work, reflection strategies such as cognitive forcing and prompts that foster critical engagement have shown promise in reducing overreliance and improving decision quality. However, these strategies have primarily been evaluated as short-term interventions within single sessions. The next challenge is to assess whether such mechanisms sustain human agency and expertise over time. Drawing on prior work in AI-assisted decision-making, metacognition, and reflective AI engagement, we examine the challenges of designing and evaluating reflective mechanisms for long-term skill sustainability, considering individual differences in how users engage with such support, the organisational conditions under which it is implemented, and the gap between short-term evidence and long-term claims. We introduce open questions for the research community about the conditions under which reflective AI engagement can be sustained in practice.
\end{abstract}




\maketitle

\section{Introduction}
Research on AI-supported professional work has extensively studied AI-assisted decision-making in single-sessions, focusing on how AI suggestions and explanations can improve human-AI decision-making~\cite{Lai2019HumanPredictionsExplanations, Zhang2020EffectConfidenceExplanation, Lai2023ScienceHumanAIDecision, Wang2021AreExplanationsHelpful}. However, these strategies have primarily been evaluated as short-term interventions for improving task performance, trust calibration, or appropriate reliance. Moreover, in these sessions, many studies did not achieve complementary AI assistance, in which the human-AI team outperforms each acting individually~\cite{Bansal2021DoesWholeExceed, Vaccaro2024WhenCombinationsHumans}. One of the underlying reasons is that people do not rely appropriately on AI assistance~\cite{Lee2004TrustAutomationDesigning}. Strategies such as pre-commitment, uncertainty communication, and added reflection steps through cognitive forcing have been shown to mitigate overreliance and improve decision quality~\cite{Vasconcelos2023ExplanationsCanReduce, Bucinca2021TrustThinkCognitive,deJong2025CognitiveForcingBetter, Green2019PrinciplesLimitsAlgorithmintheLoop, Park2019SlowAlgorithmImproves, deJong2025AlgorithmAppreciationAversion}. As these strategies tend to require additional effort from workers, the challenge is to sustain this level of reflective engagement in real-world workflows.

More recently, LLM-generated rationales have been explored to help users understand and reflect on AI decisions~\cite{Eigner2024DeterminantsLLMassistedDecisionMaking, Kim2025FosteringAppropriateReliance, Li2025TextTrustEmpowering}. However, their fluency can make incorrect outputs appear credible~\cite{Salvi2025ConversationalPersuasivenessGPT4, Breum2024PersuasivePowerLarge}, which may discourage the critical engagement they were meant to support. In group settings, the persuasiveness of LLMs may amplify majority viewpoints and increase conformity effects~\cite{deJong2025ImpactAgentGeneratedRationales}. In contrast, when effectively used, LLMs can durably reduce belief in conspiracy theories~\cite{Costello2024DurablyReducingConspiracy} or help people understand and find common ground across opposing views~\cite{Tessler2024AICanHelp, Argyle2023LeveragingAIDemocratic}. These dynamics suggest that the influence of LLM-generated rationales extends beyond individual decision-making. In group interactions, AI-mediated persuasion may shape shared professional norms within organisations.

Whether at the level of individual reasoning or collective deliberation, these findings raise a deeper concern. If the mechanisms designed to support human oversight can also undermine it, their value cannot be assessed solely by immediate decision quality. The question is whether professionals who rely on reflection mechanisms through human-AI interaction still develop their own expertise or slowly lose the capacity for independent judgement. For example, clinicians who routinely review AI-suggested diagnoses may perform well in these sessions, but are they still developing the diagnostic reasoning that unassisted practice would have demanded? A recent literature review of AI-induced deskilling in healthcare raises similar concerns across clinical specialities~\cite{Natali2025AIInducedDeskillingMedicine}, arguing that strategies to monitor and mitigate skill erosion need to be developed at both the individual workflow and organisational levels.

In this paper, we discuss four challenges for sustaining human agency and expertise through reflective human-AI interaction, and raise open questions for the research community about the conditions under which such approaches can succeed in practice.

\section{Varying Effects of Reflective Support}
Reflective interaction mechanisms assume that prompting users to reason independently will improve their engagement with AI output. Empirical evidence suggests this varies across users. Bucinca et al.\ found that cognitive forcing reduced overreliance on average, but the benefit was most prominent among participants with higher Need for Cognition (NFC)~\cite{Bucinca2021TrustThinkCognitive}. De Jong et al.\ found similar effects of NFC on cognitive forcing and additionally showed that engagement varied across task difficulties, with participants engaging with cognitive forcing when they felt capable of reasoning independently, but relying on the AI suggestions otherwise~\cite{deJong2025CognitiveForcingBetter}. This is consistent with Vasconcelos et al., who found that users weigh the effort of processing an explanation against its perceived benefit~\cite{Vasconcelos2023ExplanationsCanReduce}. Overall, these findings suggest that effectiveness does not only depend on the design of the interaction but also on the user's expertise, domain familiarity, and personal characteristics.

In practice, these differences may complicate the design of decision support and raise questions about professional autonomy. If a reflective mechanism is perceived as ineffective or unnecessary, professionals may simply ignore it. A novice professional who is still building foundational knowledge may benefit from a mechanism that withholds AI suggestions, as it allows them to develop independent reasoning. However, an expert may be able to directly evaluate AI advice critically, making withholding feel tedious instead of productive. One approach is to make reflective scaffolding adaptive. However, adaptation requires the system to assess the user's competence, which reduces the professional's autonomy over their own development. Over-scaffolding may unnecessarily constrain experienced professionals, while under-scaffolding may fail to support those still developing independent judgement. 

\textbf{Q1: When does reflective scaffolding undermine autonomy?} Mechanisms that prompt independent reasoning assume the user requires support to think critically. For experienced professionals, this assumption may be unfounded and potentially undermining. How can reflective mechanisms support skill sustainability without implying that the user lacks competence? Who decides which users receive reflective scaffolding: the system, the organisation, or the professionals themselves?

\section{Reflection in Efficiency-Driven Workflows}
A primary driver of AI adoption in professional work is to enable faster decision-making and higher throughput while reducing cognitive load. Reflective interaction mechanisms, by design, contradict this promise. They introduce friction, ask users to slow down, and demand cognitive effort at moments where the AI could have provided an immediate answer~\cite{Park2019SlowAlgorithmImproves, Bucinca2021TrustThinkCognitive, deJong2025CognitiveForcingBetter, Green2019PrinciplesLimitsAlgorithmintheLoop}. A controlled study can accommodate this trade-off. However, it is more problematic in a real-world environment that measures performance by quantity or speed.

If organisations do not recognise reflective engagement as an investment in long-term competence, these mechanisms are unlikely to be sustained in practice. Professionals using AI decision support typically need to reach productivity targets that leave little room for deliberate reflection. When reflective interactions add time to each decision and a professional makes numerous decisions per day, the cost becomes difficult to justify. This tension is particularly evident in high-stakes, risk-averse domains such as healthcare, where reflection and skill retention are closely linked to patient safety. This challenge extends beyond individual workplaces. When competing tools offer the same AI support without the added friction, organisations face pressure to adopt the faster alternative, making skill-sustaining designs harder to justify in the market.

\textbf{Q2: How can skill-sustaining reliance withstand organisational and competitive pressure?} Reflective interaction mechanisms take time. In workplaces that optimise for throughput, tools that deliberately slow professionals down may be difficult to justify, even if they preserve long-term competence. This challenge extends to the competitive landscape. If rival tools offer equivalent AI support that prioritises speed over reflection, skill-sustaining designs become a competitive disadvantage. Is this a challenge for interaction design, organisational policy, governmental policy, or a combination of these?

\section{Sustaining Independent Metacognition}
Reflective interaction mechanisms are designed to keep users cognitively engaged with AI output. For example, Fischer et al.\ describe a taxonomy of Socratic questions to promote critical reflection in AI-assisted clinical decision-making~\cite{Fischer2025TaxonomyQuestionsCritical}. The need for such prompts is supported by Fernandes et al., who found that LLMs improved task performance but made users overconfident about their own accuracy~\cite{Fernandes2026AIMakesYoua}. Most participants relied on single prompts without further follow-up, suggesting that the LLM interaction did not encourage reflection. Rather than supporting metacognition, the AI replaced the opportunity to develop it naturally. Steyvers and Peters argue more broadly that LLMs impose substantial metacognitive demands on users but are reluctant to express uncertainty~\cite{Steyvers2025MetacognitionUncertaintyCommunication}. Because humans rely heavily on linguistic cues of uncertainty, the absence of doubt can lead to unwarranted reliance, further weakening the user's capacity for self-monitoring.

Reflective scaffolding, such as structured questions or prompted reflection, may address this gap in the short term. However, when the same prompts are encountered repeatedly, their effect may wear off through habituation as users learn to respond to the prompt without critically engaging with it. This is consistent with broader findings on warning fatigue, where repeated alerts are automatically dismissed~\cite{Anderson2015WarningFatigue}. This may cause scaffolding itself to become the source of engagement rather than the user's own metacognitive habits. A professional who only reflects on AI output when prompted by the system is not necessarily developing the capacity to reflect without that prompt. Over time, the scaffolding may replace the metacognitive capacity it was meant to protect. Removing or changing the tool leaves the professional without both AI support and the self-monitoring habits that independent practice would have developed.

\textbf{Q3: Can skill-sustaining reliance itself become a form of deskilling?} If a professional relies on an AI tool's reflective prompts to maintain their reasoning practice, the mechanism may replace the professional's own metacognitive habits rather than strengthen them.

\section{From Short-Term Evidence to Long-Term Claims}

Q3 addresses a limitation that extends beyond reflective scaffolding. Most empirical work on AI-assisted decision-making evaluates interventions within single sessions, making it impossible to determine whether their effects persist. Metrics such as task accuracy, reliance rate, and trust calibration are measured immediately after exposure. Therefore, the conclusions about reflection mentioned in the introduction only address session-level consequences.

A growing body of work has begun to examine how trust and reliance develop over repeated interactions. Kahr et al. studied trust development across repeated legal decision-making tasks, showing that trust is dynamic, shaped by AI accuracy, explanation type, and the timing of errors~\cite{Kahr2023ItSeemsSmart, Kahr2024UnderstandingTrustReliance}. Riedl and Bogert tracked chess players' use of AI feedback over time, finding that how people chose to engage with AI feedback determined whether they improved or stagnated~\cite{Riedl2026WhoBenefitsAI}. In a field study with logistics professionals, Kahr et al.\ found that experts developed trust over time despite recognising AI imperfections, suggesting that apart from performance, long-term trust also depends on transparency, system consistency, and professionals' interpersonal values~\cite{Kahr2025GoodPerformanceIsnt}. These studies are an important step towards longitudinal evaluation. However, they focus on whether users appropriately rely on the AI over time, not on whether their capacity for independent reasoning is maintained or reduced through repeated interaction.

This distinction matters because the most relevant outcomes for skill sustainability may not be visible in trust or reliance measures. As discussed earlier, AI support can improve task performance while reducing users' ability to judge their own accuracy~\cite{Fernandes2026AIMakesYoua}. Similarly, Goh et al.\ found that giving physicians access to an LLM did not improve their diagnostic reasoning, even though the LLM outperformed them~\cite{Goh2024InfluenceLargeLanguage}, suggesting that the interaction did not translate AI capability into better human reasoning. If professionals can use a highly capable AI without improving their own judgement, it raises the question of what happens to that judgement over months of repeated use.

\textbf{Q4: How to navigate the asymmetry between short-term evidence and long-term claims?} Reflective mechanisms have been shown to activate reasoning in short-term controlled sessions. However, it is unclear whether they sustain expertise longitudinally. How should the field act on promising short-term findings when the long-term evidence is insufficient? What level of evidence is sufficient to justify deploying reflective mechanisms in professional practice, and what level is needed to justify deciding against their deployment?

\section{Conclusion}
The four questions raised in this paper share a common underlying challenge. Reflective AI engagement is a promising direction, but its long-term viability depends on conditions that extend beyond interaction design. Whether reflective mechanisms sustain expertise depends on who uses them, whether organisations support them, and whether they build lasting capacity. We hope these questions serve as a starting point for discussion on the sustainability of human agency and expertise in AI-supported work environments.

\bibliographystyle{ACM-Reference-Format}
\bibliography{references}

\end{document}